\documentclass[preprint,amsmath,amssymb,nofootinbib]{revtex4}
\usepackage[dvips]{graphicx}
\usepackage{rotating}
\usepackage{hyperref}
\usepackage{subfig}
\usepackage{float}
\usepackage{array}
\usepackage[utf8]{inputenc}
\usepackage[normalem]{ulem}
\usepackage{color}
\definecolor{orange}{rgb}{1,0.5,0}
\usepackage[T1]{fontenc}

\newcommand{\tvect}[2]{\ensuremath{\Bigl(\negthinspace\begin{smallmatrix}#1\\#2\end{smallmatrix}\Bigr)}}

\begin{document}

\title{Final taus and initial state polarization signatures from effective interactions of Majorana neutrinos at future $e^{+}e^{-}$ colliders.}

\author{Luc\'{\i}a Duarte}
\email{lduarte@fing.edu.uy}
 \affiliation{Instituto de F\'{\i}sica, Facultad de Ingenier\'{\i}a,
 Universidad de la Rep\'ublica \\ Julio Herrera y Reissig 565,(11300) 
Montevideo, Uruguay.}

\author{Gabriel Zapata}
\author{Oscar A. Sampayo}
\email{sampayo@mdp.edu.ar}

 \affiliation{Instituto de Investigaciones F\'{\i}sicas de Mar del Plata 
(IFIMAR)\\ CONICET, UNMDP\\ Departamento de F\'{\i}sica,
Universidad Nacional de Mar del Plata \\
Funes 3350, (7600) Mar del Plata, Argentina.}

\begin{abstract}
We study the possibility of future $e^{+}e^{-}$ colliders to disentangle different new physics contributions to the production of heavy sterile Majorana neutrinos in the lepton number violating channel $e^{+}e^{-}\rightarrow l^{+} l^{+}+  4 jets$, with $l=e, \mu, \tau$. This is done investigating the final anti-tau polarization trails and initial beam polarization dependence of the signal on effective operators with distinct Dirac-Lorentz structure contributing to the Majorana neutrino production and decay, which parameterize new physics from a higher energy scale. We find both analyses could well disentangle possible vectorial and scalar operators contributions. 

\end{abstract}

\maketitle

\section{\bf Introduction}

Individual lepton flavors and total lepton number are strictly conserved quantities in the standard model (SM). However, neutrino oscillations evidence lepton flavor violation in the neutral lepton sector, suggesting the need to consider SM extensions capable of accounting for massive light neutrinos and lepton mixing. The incorporation of light neutrino masses is still the most compelling experimental evidence of the need to enlarge the SM electroweak sector. The extensions considering sterile right-handed neutrinos with Majorana mass terms, lead to Majorana massive states which predict the occurrence of total lepton number violation (LNV). In turn, the observation of LNV would be a clear signal of new physics, and of the existence of Majorana fermions. 

The seesaw mechanism for neutrino mass generation \cite{Minkowski:1977sc, Mohapatra:1979ia, Yanagida:1980xy, GellMann:1980vs, Schechter:1980gr}, introducing right handed sterile neutrinos $N_i$ which can have a Majorana mass term leading to Majorana massive neutrino states, could account for the observation of lepton number violating processes. However, in the simplest Type I seesaw implementations, for Yukawa couplings of order $Y \sim 1$, a Majorana mass scale of order $M_{N} \sim 10^{15} GeV$ is needed to account for a light neutrino mass compatible with the current neutrino data ($m_{\nu}\sim 0.01 ~eV$)\cite{Tanabashi:2018oca}. On the other hand, for smaller Yukawa couplings, of order $Y\sim 10^{-8}-10^{-6}$, sterile neutrinos with masses around $M_{N}\sim (1-1000) ~GeV$ could exist, but this leads to negligible neutrino mixing values $U_{lN}^2 \sim m_{\nu}/M_N \sim 10^{-14}-10^{-10}$ \cite{Cai:2017mow, Atre:2009rg}. Thus, both alternatives lead to the decoupling of the Majorana neutrinos \cite{Kersten:2007vk}. 

Recent approaches consider a toy-like model in which the SM is extended by incorporating a massive Majorana sterile fermion, assumed to have non-negligible mixings with the active states, without making any hypothesis on the neutrino mass generation mechanism \cite{Abada:2017jjx, Pascoli:2018heg}. Such a minimal SM extension leads to contributions to LNV observables which are already close, or even in conflict, with current data from meson and tau decays, for Majorana masses $M_N$ below $10~GeV$ (see \cite{Abada:2017jjx, Abada:2018nio} and references therein). So, also from the experimental point of view, the simple SM extensions which attribute LNV only to the mixing between heavy Majorana states and the active neutrinos are facing increasingly stringent constraints.      

In this scenario, the observation of lepton number violating (LNV) processes allowed by the existence of a Majorana neutrino mass term would be a sign of physics beyond the minimal seesaw mechanism \cite{delAguila:2008ir} and beyond the mere existence of sterile-active neutrino mixings.   

From the theoretical point of view, one can think of an alternative approach, and consider the Majorana neutrino interactions as originating in new physics from a higher energy scale, parameterized by a model independent effective Lagrangian \cite{delAguila:2008ir}. In this approach, we consider that the sterile $N$ interacts with the SM particles by higher dimension effective operators, and take these interactions to be dominant in comparison with the mixing with light neutrinos through the Yukawa couplings, which we neglect. In this sense we depart from the usual viewpoint in which the sterile neutrinos mixing with the standard neutrinos is assumed to govern the $N$ production and decay mechanisms \cite{Atre:2009rg, delAguila:2007qnc}. 

The effective interactions we consider here for the heavy Majorana neutrinos were early studied in \cite{delAguila:2008ir}, where the possible phenomenology of dimension 6 effective operators was introduced. The dimension 5 operators extending the low-scale Type-I seesaw were investigated in \cite{Aparici:2009fh}, and their phenomenology was addressed recently in  \cite{Ballett:2016opr, Caputo:2017pit}. Dimension 7 effective $N$ operators are studied in \cite{Bhattacharya:2015vja, Liao:2016qyd}. The collider phenomenology of the dimension 6 effective Lagrangian used in this paper has been studied by our group and others in \cite{delAguila:2008ir, Peressutti:2011kx, Peressutti:2014lka, Duarte:2014zea, Duarte:2015iba, Duarte:2016miz, Duarte:2016caz, Yue:2017mmi, Duarte:2018xst}. Recently, the predictions of the effective interactions in leptonic decays of pseudoscalar mesons have been investigated in \cite{Yue:2018hci}.   

The different operators in the effective Lagrangian, with distinct Dirac-Lorentz structure, parameterize a wide variety of UV-complete new physics models, like extended scalar and gauge sectors as the left-right symmetric model, vector and scalar leptoquarks, etc. Thus, discerning the possible contributions given by them to specific processes gives us a hint on what kind of new physics at a higher energy regime is responsible for the observed interactions. 

In \cite{Duarte:2018xst} we studied the potential of final lepton angular asymmetries and initial electron polarization observables to disentangle the possible contributions of effective operators with different Dirac-Lorentz structure to the LNV $e^- p\rightarrow l^{+} + 3 jets$ process. Now we aim to take advantage of the clean environment in electron-positron colliders and exploit initial state polarization observables to distinguish the contributions from scalar and vectorial effective interactions. Also, a same-sign final anti-taus state in the $e^{+}e^{-} \rightarrow l_i^{+} l_j^{+}+  4 jets$ channel allows to measure the final tau polarization and build observables to this end. 

Lepton number violating processes have been studied thoroughly in the context of seesaw models in colliders (for comprehensive reviews on the topic see \cite{Cai:2017mow, Deppisch:2015qwa} and references therein). Lepton colliders are very well suited for the study of Majorana neutrino interactions, as they provide clean signals, without QCD jet backgrounds. The literature using lepton colliders -in past, existing and proposed experiments like the linear ILC \cite{Behnke:2013xla} or circular colliders like the FCC-ee \cite{dEnterria:2016sca} and the CEPC \cite{CEPCStudyGroup:2018rmc}- to study the production of heavy sterile neutrinos is very extensive: recent studies of the two-unit LNV channel $e^{+}e^{-} \rightarrow l_i^{\pm} l_j^{\pm}+  4 jets$, with $l_i^{\pm}=e, ~\mu, ~\tau$, in electron-positron colliders can be found in  e.g. \cite{Zhang:2018rtr, Biswal:2017nfl, Banerjee:2015gca}, and other (not necessarily LNV) heavy sterile neutrino mediated processes as $e^{+}e^{-} \rightarrow l \nu + 2 jets$ \cite{Hernandez:2018cgc, Das:2018usr, Chakraborty:2018khw, Liao:2017jiz, Banerjee:2015gca, Antusch:2015mia, Blondel:2014bra}. The initial leptons polarization in linear $e^{+}e^{-}$ colliders has been used recently in \cite{Biswal:2017nfl} to show that the comparison of polarized and unpolarized cross-sections in the $e^{+}e^{-} \rightarrow N N$ channel for the left-right symmetric model can reveal the nature of the heavy neutrino interaction with the SM sector and probe the heavy-light neutrino mixing parameters. Also, the capability to measure final tau leptons polarization has been explored in the context of neutrino mass physics. It has been widely used to distinguish different heavy scalar mediated neutrino mass generation mechanisms as Type II seesaw and the Zee-Babu model, in which the doubly charged Higgs can couple to either left-handed or right-handed leptons (see \cite{Li:2018jns, Sugiyama:2012yw} and references therein).  

The paper is organized as follows. In Sect.\ref{sec:modelo} we introduce the effective Lagrangian formalism, present the analytical calculation of the cross section for the $e^{+}e^{-} \rightarrow l_i^{+} l_j^{+}+  4 \mathrm{j}$ channel
and review the existing constraints on the effective couplings. In Sect.\ref{sec:numerical} we calculate the vectorial and scalar operators contribution to the signal cross section for different Majorana neutrino masses $m_N$ in the range $m_W \lesssim m_N$, implementing basic trigger cuts for a benchmark ILC operating scenario with $\sqrt{s}=500 ~GeV$, and comment on possible backgrounds. 
The initial beam polarization dependence of the signal is studied in Sect.\ref{sec:ee_polarization}, while the final anti-tau polarization signatures are discussed in Sect.\ref{sec:tau_polarization}. we present our final comments and conclusions in Sect.\ref{sec:Concl}.

\section{\bf Majorana neutrino interaction model} \label{sec:modelo}

\subsection{Effective operators and Lagrangian}

The effects of the new physics involving one heavy sterile neutrino $N$ and the SM fields are parameterized by a set of effective operators $\mathcal{O}_\mathcal{J}$ satisfying the $SU(2)_L \otimes U(1)_Y$ gauge symmetry \cite{Wudka:1999ax}. 
The contribution of these operators to observable quantities is suppressed by inverse powers of the new physics scale $\Lambda$. The total Lagrangian is organized as follows:

\begin{eqnarray}\label{eq:Lagrangian}
\mathcal{L}=\mathcal{L}_{SM}+\sum_{n=5}^{\infty}\frac1{\Lambda^{n-4}}\sum_{\mathcal{J}} \alpha_{\mathcal{J}} \mathcal{O}_{\mathcal{J}}^{(n)}
\end{eqnarray}
where $n$ is the mass dimension of the operator $\mathcal{O}_{\mathcal{J}}^{(n)}$.

Note that we do not include the Type-I seesaw Lagrangian -the Majorana and Yukawa terms- giving rise to the mixing between the sterile and the standard left-handed neutrinos, which we are neglecting. In this work it is considered that the dominating new physics effects leading to the lepton number violation come from the lower dimension operators that can be generated at tree level in the unknown underlying renormalizable theory.

The dimension 5 operators in \eqref{eq:Lagrangian} were studied in detail in \cite{Aparici:2009fh}. These include the well known Weinberg operator $\mathcal{O}_{W}\sim (\bar{L}\tilde{\phi})(\phi^{\dagger}L^{c})$ \cite{Weinberg:1979sa} contributing to the light neutrino masses, and operators: $\mathcal{O}_{N\phi}\sim (\bar{N}N^{c})(\phi^{\dagger} \phi)$ contributing to the $N$ Majorana masses and giving couplings of the heavy neutrinos to the Higgs (its phenomenology for the LHC has been studied very recently in \cite{Caputo:2017pit}), and an operator $\mathcal{O}^{(5)}_{NB}\sim (\bar{N}\sigma_{\mu \nu}N^{c}) B^{\mu \nu}$ inducing magnetic moments for the heavy neutrinos, which is identically zero if we include just one sterile neutrino $N$ in the theory.
In the following, as the dimension 5 operators do not contribute to the studied processes -discarding the heavy-light neutrino mixings- we will only consider the contributions of the dimension 6 operators, following the treatment presented in \cite{delAguila:2008ir}.

We organize the effective operators in different subsets.
The first one includes operators with scalar and vector bosons (SVB),
\begin{eqnarray} \label{eq:ope1}
\mathcal{O}^{(i)}_{LN\phi}=(\phi^{\dag}\phi)(\bar L_i N \tilde{\phi}), 
\;\; \mathcal{O}_{NN\phi}=i(\phi^{\dag}D_{\mu}\phi)(\bar N \gamma^{\mu} N), 
\;\; \mathcal{O}^{(i)}_{Ne\phi}=i(\phi^T \epsilon D_{\mu} \phi)(\bar N \gamma^{\mu} e_i)
\end{eqnarray}
and a second subset includes the baryon-number conserving 4-fermion ($4-f$) contact terms:
\begin{eqnarray} \label{eq:ope2}
\mathcal{O}^{(i,j)}_{duNe}&=&(\bar d_{i} \gamma^{\mu} u_{i})(\bar N \gamma_{\mu} e_{j}) ,  
\;\; \mathcal{O}^{(i,j)}_{LNLe}=(\bar L_i N)\epsilon (\bar L_j e_j),
\nonumber \\
\mathcal{O}^{(i,j)}_{LNQd}&=&(\bar L_i N) \epsilon (\bar Q_j d_j), 
\;\; \mathcal{O}^{(i,j)}_{QuNL}=(\bar Q_i u_i)(\bar N L_j) , 
\;\; \mathcal{O}^{(i,j)}_{QNLd}=(\bar Q_i N)\epsilon (\bar L_j d_j) , 
\nonumber \\
\;\; \mathcal{O}^{(i)}_{fNN}&=&(\bar f_i \gamma^{\mu} f_i)(\bar N \gamma_{\mu}N),
\;\; \mathcal{O}^{(i)}_{LN}= |\bar{N} L_i|^2
\end{eqnarray}
where $e_i$, $u_i$, $d_i$ and $L_i$, $Q_i$ denote, for the family
labeled $i$ (or $j$), the right handed $SU(2)$ singlets and the left-handed
$SU(2)$ doublets, respectively. The symbol $f$ in the $\mathcal{O}^{(i)}_{fNN}$ operator stands for every SM fermion. Here $ \gamma^{\mu}$ are the Dirac matrices, and $\epsilon=i\sigma^{2}$ is the antisymmetric symbol. In this work we allow for family mixing, letting the family indices to be different in the operators that can involve more than one SM fermion family. 

We also consider the one-loop ($1-loop$) generated operators, which are naturally suppressed by a factor $1/16\pi^2$  \cite{delAguila:2008ir, Arzt:1994gp}. These operators give interaction terms that are involved in the full calculation of the Majorana neutrino total width $\Gamma_N$, and the branching ratios of its different decay channels. Their expressions can be found in \cite{Duarte:2016miz}. 

In order to obtain the interactions in the process $e^{+}e^{-}\rightarrow l_{i}^{+} l_{j}^{+}+ 4 \mathrm{j}$, we consider the effective Lagrangian terms involved in the calculations, taking the scalar doublet after spontaneous symmetry breaking as $\phi=\tvect{0}{\frac{v+h}{\sqrt{2}}}$, with $h$ being the Higgs field and $v$ its v.e.v. We only write here the Lagrangian terms involved in the production and decay processes considered in the current calculation. For the complete dimension 6 Lagrangian, we refer the reader to Appendix A in \cite{Duarte:2016miz}.

The operators in \eqref{eq:ope1} contribute to a first Lagrangian piece
\begin{eqnarray}\label{eq:leff_svb_C}
  \mathcal{L}^{tree}_{SVB}= && \frac{1}{\Lambda^2}\left\{\alpha_Z (\bar N_R \gamma^{\mu} N_R)  \left( \frac{ m_Z ~v}{2} Z_{\mu} - \frac{v}{2} P^{(h)}_{\mu} h + ...  \right) \right.
\nonumber
\\ && \left. -\alpha^{(i)}_W (\bar N_R \gamma^{\mu} e_{R,i})\left(\frac{m_{W} ~v }{\sqrt{2}}W^{+}_{\mu} + ... \right) + h.c. \right\}.
\end{eqnarray}
and the 4-fermion interactions involving quarks and leptons from \eqref{eq:ope2} give
\begin{eqnarray}\label{eq:leff_4-f}
\mathcal{L}^{tree}_{4-f}&=& \frac{1}{\Lambda^2} \left\{ \alpha^{(i,j)}_{V_0} \bar d_{R,j} \gamma^{\mu} u_{R,j} \bar N_R \gamma_{\mu} e_{R,i} + \alpha^{(i)}_{V_1} \bar e_{R,i} \gamma^{\mu} e_{R,i} \bar N_R \gamma_{\mu} N_R + \alpha^{(i)}_{V_2} \bar L_i \gamma^{\mu} L_i \bar N_R \gamma_{\mu} N_R + \right. \nonumber
\\ && \left.
\alpha^{(i,j)}_{S_0}(\bar \nu_{L,j}N_R \bar e_{L,i}e_{R,i}-\bar e_{L,j}N_R \bar \nu_{L,i}e_{R,i}) + \alpha^{(i,j)}_{S_1}(\bar
u_{L,j}u_{R,j}\bar N_R \nu_{L,i}+\bar d_{L,j}u_{R,j} \bar N_R e_{L,i})
 + \right. \nonumber
\\ && \left.
\alpha^{(i,j)}_{S_2} (\bar \nu_{L,i}N_R \bar d_{L,j}d_{R,j}-\bar e_{L,i}N_R \bar u_{L,j}d_{R,j}) + \alpha^{(i,j)}_{S_3}(\bar
u_{L,j}N_R \bar e_{L,i}d_{R,i}-\bar d_{L,j}N_R \bar \nu_{L,i}d_{R,i}) + \right. \nonumber
\\ && \left.  \alpha^{(i)}_{S_4} (\bar N_R \nu_{L,i}~\bar \nu_{L,i} N_R~+\bar
N_R e_{L,i} \bar e_{L,i} N_R) + \cdots  + h.c. \right\}.
\end{eqnarray}
In Eqs. \eqref{eq:leff_svb_C} and \eqref{eq:leff_4-f} a sum over the family index $i,j=1,2,3$ is understood, and the couplings
$\alpha^{(i,j)}_{\mathcal O}$ are associated to specific operators:
\begin{eqnarray}\label{eq:alphas_nombre}
\alpha_Z&=&\alpha_{NN\phi},\; 
\alpha^{(i)}_W=\alpha^{(i)}_{Ne\phi},\;
\alpha^{(i,j)}_{V_0}=\alpha^{(i,j)}_{duNe},\;\;
\alpha^{(i)}_{V_1}=\alpha^{(i)}_{eNN},\;\; 
\alpha^{(i)}_{V_2}=\alpha^{(i)}_{LNN} \nonumber \\
\alpha^{(i,j)}_{S_0}&=&\alpha^{(i,j)}_{LNLe},\;\;   
\alpha^{(i,j)}_{S_1}=\alpha^{(i,j)}_{QuNL},\; \alpha^{(i,j)}_{S_2}=\alpha^{(i,j)}_{LNQd},\;\;
\alpha^{(i,j)}_{S_3}=\alpha^{(i,j)}_{QNLd},\; \alpha^{(i)}_{S_4}=\alpha^{(i)}_{LN}.
\end{eqnarray}

The effective operators above can be classified by their Dirac-Lorentz structure into {\it{scalar}}, {\it{vectorial}} and {\it{tensorial}}. The scalar and vectorial operators contributing to the studied processes are those appearing in \eqref{eq:leff_svb_C} and \eqref{eq:leff_4-f} with couplings named $\alpha_{S}$ and $\alpha_{W, ~Z, ~ V}$, respectively. For the Majorana neutrinos production vertices, depicted in Figs.\ref{fig:eeNN} and \ref{fig:eeNW}, and the decay process $N\rightarrow l^{+} \mathrm{j}\mathrm{j}$ in Fig.\ref{fig:Nlep}, we have scalar and vectorial contributions from the effective Lagrangian related to the spontaneous symmetry breaking process coming from \eqref{eq:ope1} and the 4-fermion interactions involving quarks and leptons from \eqref{eq:ope2}. The dimension 6 tensorial operators are generated at one-loop level, and they are suppressed by the loop factor $1/16\pi^2$ with respect to the considered operators. They do take part in the calculation of the total width $\Gamma_N$. The relative sizes between the different effective couplings are given by the contribution of the corresponding operators to the experimental observables.

\subsection{Signal}\label{sec:signal}

In this work we study the possibility for future $e^{+}e^{-}$ colliders to produce clear signatures of Majorana
neutrinos in the context of interactions coming from an effective Lagrangian approach in the $e^{+}e^{-}\rightarrow l_{i}^{+} l_{j}^{+}+  4 \mathrm{j}$ process. 

\begin{figure*}[h]
\centering
 \includegraphics[width=0.8\textwidth]{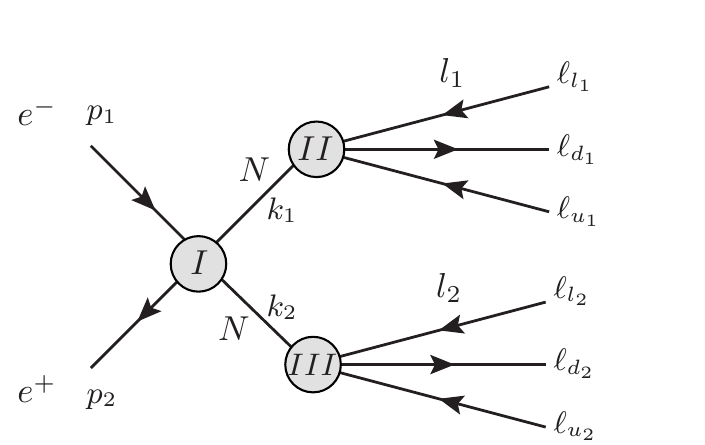}
\caption{\label{fig:eeNN} Diagrams contributing to double $N$ production.}
\end{figure*}

\begin{figure*}[h]
\centering
 \includegraphics[width=0.8\textwidth]{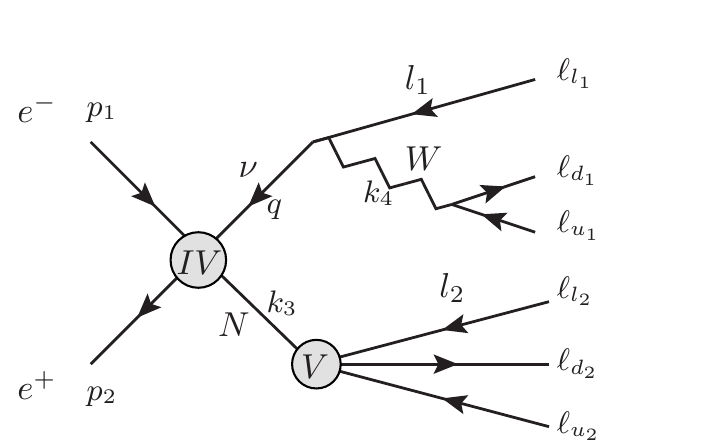}
\caption{\label{fig:eeNW} Diagrams contributing to single $N$ production.}
\end{figure*}

\begin{figure*}[h]
\centering
 \includegraphics[width=0.8\textwidth]{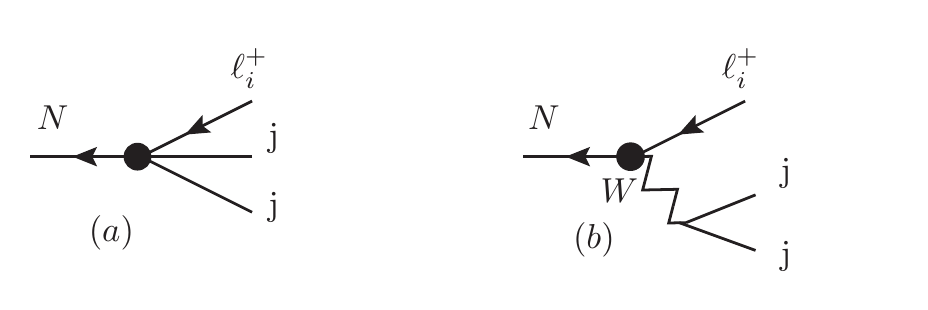}
\caption{\label{fig:Nlep} Diagrams contributing to the decay process $N \rightarrow \ell^{+}_{i} \mathrm{j}\mathrm{j}$.  }
\end{figure*}

In particular, here we show the calculation for the reaction with final anti-taus $e^{+}e^{-}\rightarrow \tau_{1}^{+}\tau_{2}^{+} + 4 \mathrm{j}$, which is divided into two subprocesses depicted in Figs. \ref{fig:eeNN} and \ref{fig:eeNW}. In the first case we consider the production of two Majorana neutrinos $N$ which will decay into one anti-tau and jets $N\rightarrow \tau^{+} \mathrm{j}\mathrm{j}$ as in Fig.\ref{fig:Nlep}. In the second case, we consider the production of a single Majorana neutrino, with the same decay as before, and a $W$ decaying into two jets $W\rightarrow \mathrm{j} \mathrm{j} $.

The differential cross section for the process in Fig. \ref{fig:eeNN} can be decomposed as a product:

\begin{eqnarray}\label{eq:dsigNN}
 d \sigma_{NN}= \frac{1}{8 ~s ~m^2_N ~\Gamma^2_N} && |\overline{M_{I}}|^2 \left[ (2\pi)^4 \delta^4(p_1+p_2-k_1-k_2) \delta(k_1^2-m^2_N) \delta(k_2^2-m^2_N)  \frac{d^4 k_1}{(2\pi)^3} \frac{d^4 k_2}{(2\pi)^3} \right]
\nonumber
\\ && 
 |\overline{M_{II}}|^2 \left[ (2\pi)^4 \delta^4\left( k_1- \sum_{i={l_1},{d_1},{u_1} } \ell_i\right)  \prod_{i={l_1},{d_1},{u_1} } \delta(\ell^2_i-m^2_i) \frac{d^4 \ell_i}{(2\pi)^3}  \right]
 \nonumber
\\ &&
 |\overline{M_{III}}|^2 \left[ (2\pi)^4 \delta^4\left( k_2- \sum_{j={l_2},{d_2},{u_2} } \ell_j\right)  \prod_{j={l_2},{d_2},{u_2} } \delta(\ell^2_j-m^2_j) \frac{d^4 \ell_j}{(2\pi)^3}  \right].
\end{eqnarray}

\begin{figure*}[h]
\centering
 \includegraphics[width=0.8\textwidth]{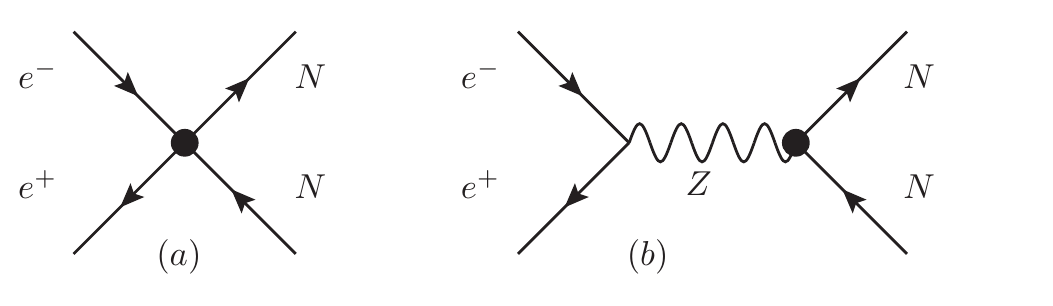}
\caption{\label{fig:Nee} Diagrams contributing to double $N$ production.}
\end{figure*}
The $NN$ production squared amplitude $|\overline{M_{I}}|^2$ involves the effective and standard $Z$ interactions in Fig. \ref{fig:Nee}. It can be written as
\begin{eqnarray}\label{eq:MI}
 |\overline{M_{I}}|^2=  \frac{1}{4} \frac{1}{\Lambda^4}\left[ 4 (\alpha^{(1)}_{S_4}+ 2 \alpha_2)^2 (p_1.k_1) (p_2.k_2)+ 16 ~\alpha^2_{1} (p_1.k_2) (p_2.k_1) \right]
\end{eqnarray}
with $\alpha^{(1)}_{S_4}$ the 4-fermion $LN$ scalar coupling in \eqref{eq:leff_4-f} and the vector combinations
\begin{eqnarray}
 \alpha_1 &&= \alpha_Z ~\Pi_Z  ~g_R + \alpha^{(1)}_{V_1} 
 \nonumber
\\ 
 \alpha_2 &&= \alpha_Z ~\Pi_Z ~g_L + \alpha^{(1)}_{V_2}. 
\end{eqnarray}
Here the $Z$ boson propagator is $\Pi_Z = \left(\frac{m^4_Z}{((p_1+p_2)^2 -m^2_Z)^2+ m^2_Z \Gamma^2_Z}\right)^{\frac{1}{2}}$, $g_R=\sin^{2}(\theta_W)$ and $g_L=-1/2 + \sin^{2}(\theta_W)$ are the SM couplings of the $Z$ boson in the initial vertex in Fig.\ref{fig:Nee} (b). We neglect the contribution of a Higgs mediated diagram similar to Fig.\ref{fig:Nee} (b), as it scales like $(\frac{m_e}{v})^2$.

The differential cross section for the single $N$ process in Fig. \ref{fig:eeNW} can be decomposed as a product:

\begin{eqnarray}\label{eq:dsigNW}
  &&  d \sigma_{NW} = \frac{1}{8 ~s ~m_N ~m_W ~\Gamma_N ~\Gamma_W}  
\nonumber
\\ && |\overline{M_{IV}}|^2
\left[ (2\pi)^4 \delta^4(p_1+p_2-k_3-k_4-\ell_{l_1}) \delta(k_3^2-m^2_N) \delta(k_4^2-m^2_W) \delta(\ell_{l_1}^2-m^2_{l_1}) \frac{d^4 k_3}{(2\pi)^3} \frac{d^4 k_4}{(2\pi)^3} \frac{d^4 \ell_{l_1}}{(2\pi)^3} \right]
\nonumber
\\ &&
 |\overline{M_{V}}|^2 \left[ (2\pi)^4 \delta^4\left( k_3- \sum_{j={l_2},{d_2},{u_2} } \ell_j\right)  \prod_{j={l_2},{d_2},{u_2} } \delta(\ell^2_j-m^2_j) \frac{d^4 \ell_j}{(2\pi)^3}  \right]
 \nonumber
\\ &&
|\overline{M_{VI}}|^2 \left[ (2\pi)^4 \delta^4\left( k_4- \sum_{i={d_1},{u_1}} \ell_i \right)  \prod_{i={d_1},{u_1} } \delta(\ell^2_i-m^2_i) \frac{d^4 \ell_i}{(2\pi)^3}  \right].
\end{eqnarray}

The $e^{-} e^{+} \rightarrow N \nu$ production amplitude in Fig. \ref{fig:eeNW} is governed by the scalar 4-fermion interaction $LNLe$:
\begin{eqnarray}\label{eq:M_IV}
 |\overline{M_{IV}}|^2 =  \frac{g^2}{2 \Lambda^4} \frac{1}{(q^2)^2}~ {\alpha^{(1,3)}_{S_0}}^2 ~\frac{(k_3.p_2)}{m^2_W} 
 && \left[ 4 (k_4.\ell_{l_1}) (k_4. q) (p_1. q) + 2 (\ell_{l_1}.q) (p_1. q) m^2_W 
  \right.
 \nonumber
\\  && \left. 
 - 2 (k_4. \ell_{l_1}) (k_4.p_1)  q^2 - m^2_W (\ell_{l_1}.p_1) q^2 \right]
\end{eqnarray}

where $q^2= (k_4+ \ell_{l_1})^2=m^2_W + 2 (k_4.\ell_{l_1})$ is the squared momentum of the neutrino, and $g$ is the SM $SU(2)_L$ coupling \footnote{In the case of final positrons $e^{-} e^{+} \rightarrow e^{+} e^{+} + 4 \mathrm{j}$, there is another diagram with a contribution proportional to ${\alpha^{(1)}_W}^2$ to the amplitude $M_{IV}$ in \eqref{eq:M_IV}. It is included in our numerical calculations.}.

The amplitudes $|\overline{M_{x}}|^2$, with $x=II, III, V$ in \eqref{eq:dsigNN} and \eqref{eq:dsigNW} represent the $N$ decay process into an anti-lepton and jets $N\rightarrow l^{+} \mathrm{j}\mathrm{j}$ depicted in Fig.\ref{fig:Nlep}. They can be written as:
\begin{eqnarray}\label{eq:M_Ndec}
 &&|\overline{M_{x}}|^2= \left(|\Lambda^{L}_{x}|^2+ |\Lambda^{R}_{x}|^2\right)
  \nonumber
\\ &&
|\Lambda^{L}_{x}|^2=\frac{16}{\Lambda^{4}}\left[ \Pi^{2}_{W} {\alpha^{(3)}_W}^2 (k_N. \ell_{u_x})(\ell_{l_x}. \ell_{d_x})+ {\alpha^{(3,j)}_{V_0}}^2 (k_N. \ell_{d_x}) (\ell_{l_x}. \ell_{u_x}) \right]
  \nonumber
\\ &&
|\Lambda^{R}_{x}|^2=\frac{4}{\Lambda^{4}}\left[ ({\alpha^{(3,j)}_{S_1}}^2 + {\alpha^{(3,j)}_{S_2}}^2- {\alpha^{(3,j)}_{S_2}}{\alpha^{(3,j)}_{S_3}})  (\ell_{d_x}. \ell_{u_x})(\ell_{l_x}. k_N)+ 
\right.
\nonumber
\\ && \left.
({\alpha^{(3,j)}_{S_3}}^2- {\alpha^{(3,j)}_{S_2}}{\alpha^{(3,j)}_{S_3}}) (\ell_{l_x}. \ell_{d_x}) (k_N. \ell_{u_x}) + {\alpha^{(3,j)}_{S_2}}{\alpha^{(3,j)}_{S_3}} (\ell_{l_x}. \ell_{u_x}) (k_N. \ell_{d_x})\right]. 
\end{eqnarray}
Here $k_N$ corresponds in each case to the momentum of the $N: k_1, k_2, k_3$ for $x= II, III, V$, as indicated in Figs. \ref{fig:eeNN} and \ref{fig:eeNW}, and the index $j=1,2$ corresponds to the final quarks family. The $W$ boson propagator is $\Pi_W = \left(\frac{m^4_W}{((k_N- \ell_{l_x})^{2} -m^2_W)^2+ m^2_W \Gamma^2_W}\right)^{\frac{1}{2}}$. 

In fact, as we are summing over the light-quarks in the final state ($u,d,c,s$), the contributions from these decays can be written as
\begin{eqnarray}
 |\Lambda^{L}_{x}|^2 &&=\frac{16}{\Lambda^{4}}\left[ \Pi^{2}_{W} C_0 (k_N. \ell_{u_x})(\ell_{l_x}. \ell_{d_x})+ C_1 (k_N. \ell_{d_x}) (\ell_{l_x}. \ell_{u_x}) \right]
  \nonumber
\\ 
|\Lambda^{R}_{x}|^2 && =\frac{4}{\Lambda^{4}}\left[ C_2  (\ell_{d_x}. \ell_{u_x})(\ell_{l_x}. k_N)+ C_3 (\ell_{l_x}. \ell_{d_x}) (k_N. \ell_{u_x}) + C_4 (\ell_{l_x}. \ell_{u_x}) (k_N. \ell_{d_x})\right]
\end{eqnarray}
where
\begin{eqnarray}\label{eq:Ces}
&&  C_0= 2 {\alpha^{(3)}_W}^2,  \qquad C_1= \sum_{j=1,2} {\alpha^{(3,j)}_{V_0}}^2, \qquad C_2= \sum_{j=1,2}({\alpha^{(3,j)}_{S_1}}^2+ {\alpha^{(3,j)}_{S_2}}^2 - {\alpha^{(3,j)}_{S_2}}{\alpha^{(3,j)}_{S_3}}) 
   \nonumber
\\ &&
C_3= \sum_{j=1,2}({\alpha^{(3,j)}_{S_3}}^2- {\alpha^{(3,j)}_{S_2}} {\alpha^{(3,j)}_{S_3}}), \qquad C_4 = \sum_{j=1,2}({\alpha^{(3,j)}_{S_2}} {\alpha^{(3,j)}_{S_3}}).
\end{eqnarray}
Each term $|\Lambda^{R,L}_{x}|^2$ in \eqref{eq:M_Ndec} gives the contribution of a $\pm$ polarized final anti-tau. We can clearly see here that the vectorial operators in $C_0$ and $C_1$ will give a contribution to Left-polarized $(-)$ final anti-taus, and the scalar operators in $C_2$, $C_3$ and $C_4$ will contribute to Right-polarized $(+)$ anti-taus.  

The amplitudes in \eqref{eq:M_Ndec} are proportional to the Majorana neutrino mass, which is the only source of LNV. This can be seen by taking into account that these are Lorentz invariant expressions. When one considers the Majorana $N$ in its rest frame, the dot products of $k_N$ with the final momenta $\ell$ ($\ell_{u_x}, \ell_{d_x}, \ell_{l_x}$) are proportional to $m_N$.

The amplitude $|\overline{M_{VI}}|^2$ in \eqref{eq:dsigNW} represents the standard decay of the $W$ boson into two light-quark ($u, d, c, s$) jets:
\begin{eqnarray}
 |\overline{M_{VI}}|^2= 2 (2 g^2 (\ell_d. \ell_u)) .
\end{eqnarray}

As we already mentioned, the total decay width of the Majorana neutrino $\Gamma_N$ appearing in the denominators in Eqs. \eqref{eq:dsigNN} and \eqref{eq:dsigNW} is calculated considering all the possible decay channels, as in \cite{Duarte:2016miz}. 

\subsection{Bounds on the effective couplings}\label{subsec:bounds}

The dimensionless effective couplings $\alpha_\mathcal{J}$ associated to the distinct operators in the Lagrangian weight the contribution of the interactions parameterized by each operator. We can divide them into two groups: those which correspond to operators involving only one heavy Majorana neutrino $N$ ($\alpha_N \equiv \alpha^{(i)}_W, \alpha^{(i,j)}_{V_0}, \alpha^{(i,j)}_{S_0}, \alpha^{(i,j)}_{S_1}, \alpha^{(i,j)}_{S_2}, \alpha^{(i,j)}_{S_3}$) and those involving two $N$s ($ \alpha_{NN}\equiv \alpha_Z, \alpha^{(i)}_{V_1}, \alpha^{(i)}_{V_2}, \alpha^{(i)}_{S_4}$)  in \eqref{eq:alphas_nombre}. The first group of couplings $\alpha_N$, for each lepton family $i,j=1,2,3$ appear in the $N$ decays in Fig. \ref{fig:Nlep} and/or in the total decay width $\Gamma_N$ \cite{Duarte:2016miz}, while the second group $\alpha_{NN}$ contribute in the double $N$ production process in Fig. \ref{fig:eeNN}. 

The numerical value of the couplings $\alpha_{N}$ can be constrained exploiting the current experimental bounds on the light-heavy neutrino mixing parameters in seesaw models. In the literature \cite{Cvetic:2018elt, Das:2017nvm, Fernandez-Martinez:2016lgt, deGouvea:2015euy, Deppisch:2015qwa, Antusch:2015mia} the existing experimental bounds are summarized in general phenomenological approaches considering low scale minimal seesaw models, parameterized by a single heavy neutrino mass scale $M_{N}$ and a light-heavy mixing $U_{lN}$, with $l$ indicating the lepton flavor. In the Majorana neutrino mass region we are considering, the most stringent constraints are placed on the $N-\nu_e$ mixing $U_{eN}$ by neutrinoless double beta decay ($0\nu\beta\beta$) searches. The $N-\nu_{\mu}$ and $N-\nu_{\tau}$ mixings $U_{\mu N}$ and $U_{\tau N}$ take their most stringent bounds from lepton flavor violating radiative decays as $\mu \rightarrow e \gamma$ and $\tau \rightarrow e (\mu) \gamma$. 

We interpret the current bounds on the $U_{lN}$ seesaw mixings comparing the effective couplings $\alpha_{N}$ with the general structure usually taken for the interaction between the heavy Majorana neutrinos and the $W$:
\begin{eqnarray}
\label{eq:lW}
 \mathcal L_W = -\frac{g}{\sqrt{2}}  \overline l \gamma^{\mu} U_{lN} P_L N W_{\mu} + h.c. 
\end{eqnarray}
The term with coupling $\alpha_{W}^{(i)}$ in \eqref{eq:leff_svb_C} can be compared to the weak charged current in \eqref{eq:lW}, giving us a relation between $\alpha_{W}^{(i)}$ and $U_{l_{i}N}$ for each fermion family $i=1,2,3$: $ U_{l_{i}N}\simeq \frac{\alpha^{(i)}_{W}v^2}{2\Lambda^2}$ \cite{delAguila:2008ir}. In order to put reliable bounds on the effective couplings $\alpha_N$ but keeping the analysis as simple as possible, we consider the bounds on the seesaw mixings to constrain all the effective couplings $\alpha_{N}^{(i)}$ for each family $i$. In previous work \cite{Duarte:2016miz, Duarte:2015iba} we have presented our procedure, and refer the reader to those papers for a detailed discussion. 

For the couplings involving the first fermion family -taking indices $i=1$ and $j=1$ in the Lagrangian terms in \eqref{eq:leff_svb_C} and \eqref{eq:leff_4-f}- the most stringent are the $0\nu\beta\beta$-decay bounds obtained by the KamLAND-Zen collaboration \cite{KamLAND-Zen:2016pfg}. Following the treatment made in \cite{Mohapatra:1998ye, deGouvea:2015euy, Duarte:2016miz}, they give us an upper limit $\alpha^{b (e)}_{0\nu\beta\beta} \leq 3.2 \times 10^{-2} \left(\frac{m_N}{100 ~GeV}\right)^{1/2}$, where the new physics scale is taken to be $\Lambda=1~TeV$ (here and in the following) \footnote{The new physics scale $\Lambda=1~TeV$ is taken as an illustration. One can obtain the values at any other scale $\Lambda^{\prime}$ considering $\alpha^{\prime}_{\mathcal{J}}=(\frac{\Lambda^{\prime}}{\Lambda})^{2} \alpha_{\mathcal{J}}$. }. For the second and third fermion families -taking indices $i=2,3$ or $j=2,3$ in \eqref{eq:leff_svb_C} and \eqref{eq:leff_4-f}- and sterile neutrino masses in the range $m_W\lesssim m_N$ the upper limits come from radiative lepton flavor violating (LFV) decays as $\mu\rightarrow e \gamma$ and  $\tau \rightarrow e (\mu) \gamma$. For the second family the constraint $Br(\mu \rightarrow e \gamma)< 5.7\times 10^{-13}$ translates into a bound $\alpha^{b (\mu)}_{LFV} \leq 0.32$ and for the third, the bound $Br(\tau \rightarrow \mu \gamma)< 1.8 \times 10^{-8}$ gives us $\alpha^{b (\tau)}_{LFV} \leq 2.48$ \cite{deGouvea:2015euy, Antusch:2015mia, Cvetic:2018elt}.

The effective couplings of the operators in the second group, involving two heavy Majorana neutrinos $\alpha_{NN}$ can be bounded exploiting the LEP results on single $Z\rightarrow \nu N$ and pair $Z\rightarrow N~N$ sterile neutrino production searches \cite{Decamp:1991uy}. However, for the $m_N$ range studied in this work ($m_W \lesssim m_N$), they do not give us any restriction on the couplings. 

In the numerical analysis throughout this work we will take a very conservative approach and consider the most   possible restricting bounds: the couplings involved in neutrinoless double beta decay ($\alpha^{1}_{W}, \alpha^{(1,1)}_{V_0}, \alpha^{(1,1)}_{S_{1,2,3}}$) are taken as equal to the bound $\alpha^{b}_{0\nu\beta\beta}= 3.2 \times 10^{-2} \left(\frac{m_N}{100 ~GeV}\right)^{1/2}$ for $\Lambda=1 ~TeV$, and all the others (scalar, and vectorial, involving one or two Majorana neutrinos) will be taken as equal to the LFV bound $\alpha^{b (\mu)}_{LFV} \leq 0.32$.

All the couplings of the operators generated at one loop level (which contribute to the total width $\Gamma_{N}$) are fixed as the corresponding tree-level coupling divided by the loop factor: $\alpha_{1-loop}={\alpha^{(i)}_{N}}/{16 \pi^{2}}$.

 \section{\bf Numerical analysis}\label{sec:numerical}
 
In our numerical analysis we aim to study the possibility of distinguishing the contributions from vectorial and scalar effective interactions in the process $e^{+}e^{-}\rightarrow l^{+}l^{+}+4\mathrm{j}$, mediated by Majorana $N$ neutrinos. This signal can be studied in future lepton colliders like the linear ILC \cite{Behnke:2013xla} or circular colliders like the FCC-ee \cite{dEnterria:2016sca} and the CEPC \cite{CEPCStudyGroup:2018rmc}. 

For concreteness, throughout the paper we will consider an $e^{+}e^{-}$ collider with center of mass energy  $\sqrt{s}=500 ~GeV$ and integrated luminosity $\mathcal{L}=500 ~fb^{-1}$ for estimating the numbers of events. These values correspond to one of the proposed ILC operation modes \cite{Baer:2013cma}. We will also exploit the possibility the ILC (and other) facilities offer to use initially polarized beams and measure final state tau polarization.   

For the effective interaction model, we will consider a new physics energy scale $\Lambda=1 ~TeV$, keeping $ \alpha  s < \Lambda^2$ in order to ensure the validity of the effective Lagrangian approach \footnote{For instance, with the bounds on the effective couplings discussed in Sect.\ref{subsec:bounds}, the  EFT expansion parameter (for the second and third fermion families) is $\frac{\alpha s}{\Lambda^2}=0.08$ for the scalar and vectorial terms, and $\frac{\alpha s}{\Lambda^2}=0.0005$ for the tensorial terms, with $\Lambda =1 ~TeV$ and $\sqrt{s}= 0.5 ~TeV$.}.

In order to consider the contributions given by the scalar operators, we set the effective couplings corresponding to the vectorial operators $\alpha^{(i)}_W$, $\alpha_Z$ and $\alpha^{(i,j)}_{V_0}$ $\alpha^{(i)}_{V_{1,2}}$ and the tensorial operators (involved in the numerical calculation of the decay width $\Gamma_N$) equal to zero, and set the value of the scalar couplings $\alpha^{(i,j)}_{S_{1,2,3}}$ and  $\alpha^{(i,j)}_{S_{0}}$ in \eqref{eq:alphas_nombre} to the maximum allowed values in Sect.\ref{subsec:bounds} corresponding to each fermion family $i,j$. In the plots, the curves labeled {\it{scalar}} correspond to the numerical evaluation in which all the vectorial (and tensorial) couplings are set to zero, and all the scalar couplings are set to the value of the bound (at the same time).  Conversely, the curves labeled {\it{vectorial}}  study the contribution from the vectorial (plus the tensorial) operators, and we set the scalar couplings to zero, taking all the vectorial couplings equal to the bound in Sect.\ref{subsec:bounds}, and the tensorial ones to this value multiplied by the loop factor $1/16\pi^2$. The vectorial and tensorial operators are considered together, because they involve the interactions of the Majorana neutrinos with the standard vector bosons ($W^{\pm}$, $Z$, photons) and the Higgs. As we already mentioned, the tensorial operators (generated at one-loop level in a possible UV-complete theory and therefore suppressed by the loop factor) give their major contribution to the decay  $N\rightarrow \nu \gamma$, which is the dominant channel only for Majorana masses $m_N\lesssim 30 ~GeV$ \cite{Duarte:2016miz}, well below the Majorana neutrino mass range considered here.

\subsection{Acceptance cuts and SM background}\label{sec:cutsybkg}

For the numerical study, we calculate the cross section for the process $e^{+}e^{-}\rightarrow l^{+}l^{+}+4\mathrm{j}$
according to the production and decay channels presented in Sect. \ref{sec:signal}. The phase space integration of the squared amplitudes is made generating the final momenta with the Monte Carlo routine RAMBO \cite{Kleiss:1985gy}.  

\begin{figure*}[h]
\centering
\includegraphics[totalheight=8cm]{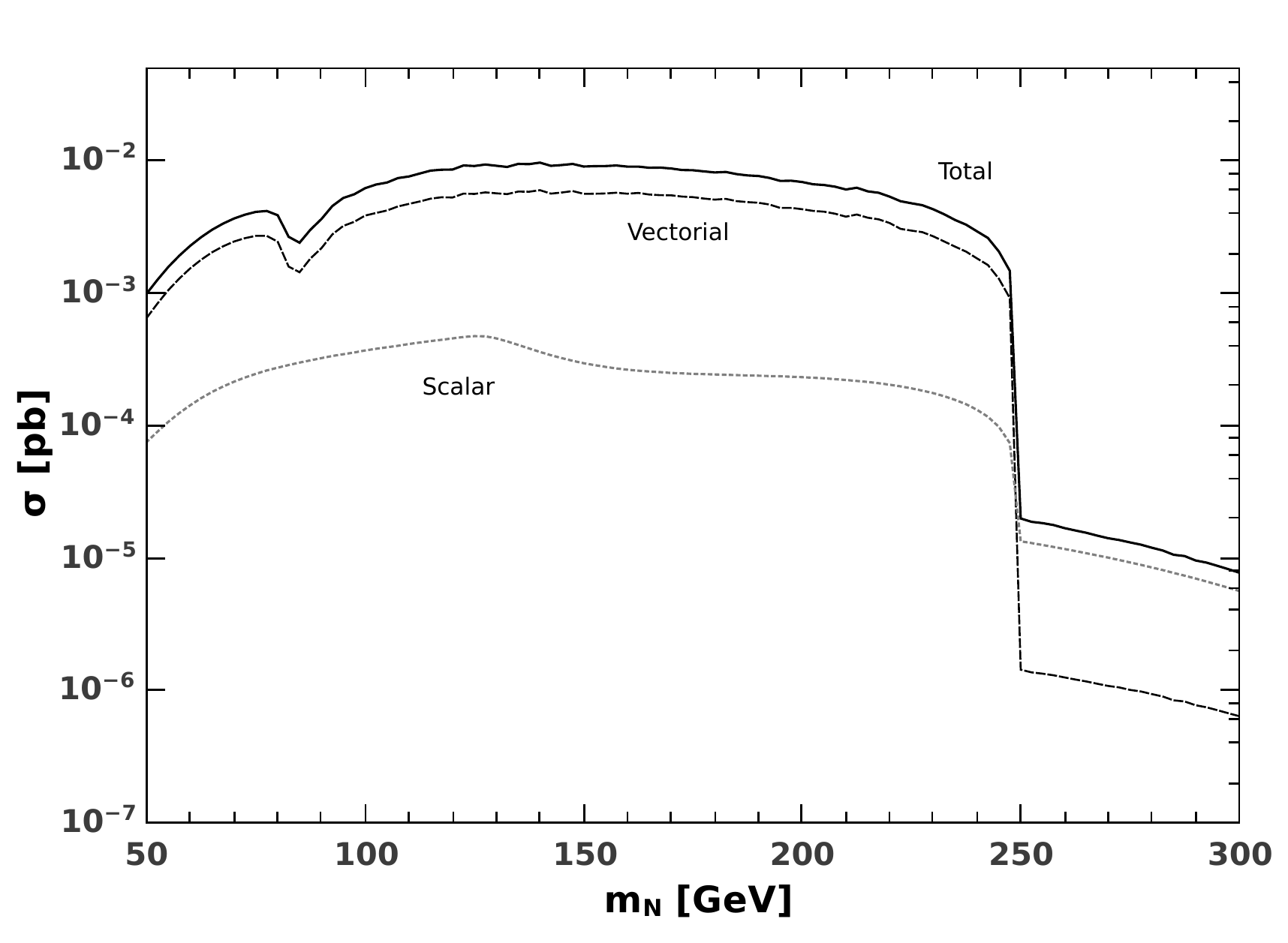}
\caption{\label{fig:sigma} Cross section for the process $e^+ e^- \rightarrow l^{+} l^{+} + 4 \mathrm{j}$}
\end{figure*}

In Fig.\ref{fig:sigma} we show the results for the signal cross section, as a function of the Majorana neutrino mass $m_N$, considering all same-sign anti-lepton final states with $l= e, \mu, \tau$. We have implemented basic trigger cuts following the generic ILC detector design \cite{Behnke:2013xla}, taking $p_T^l>10 ~GeV$ and $\lvert \eta_l \rvert<2.5$ for the final leptons, $p_T^{\mathrm{j}}>20 ~GeV$ and $\lvert \eta_{\mathrm{j}} \rvert <5$ for the jets, and a separation $\Delta R_{\mathrm{j} \, \mathrm{j}}, \Delta R_{l\,\mathrm{j}}>0.4$ between the final leptons and jets.

It can be appreciated that as $m_N$ approaches the c.m. energy limit, the cross section drops sharply. The vectorial operators give a greater contribution to the unpolarized cross section by nearly one order of magnitude. This behavior was previously found for other effective $N$ interaction signals studied in the past \cite{Duarte:2018xst, Duarte:2016caz}.

The studied signal, being a LNV process, is strictly forbidden in the SM, and it is a clean signal with practically no SM background, which appears to be mainly due to charge misidentification of one of the final leptons.
In the case of final electrons, the signal $e^{+}e^{+}+ 4\mathrm{j}$ can be faked by genuine opposite sign electron SM events. This charge-flip events are final $e^{+}e^{-}+ 4\mathrm{j}$ events in which an $e^{-}$ undergoes bremsstrahlung in the tracker volume and the associated photon decays into an $e^{+}e^{-}$ pair, and this $e^{+}$ is mistaken for the primary $e^{-}$ if it carries a large fraction of the original energy. This effect is negligible for final muons and taus. When considering electrons in the final state ($e^{+}e^{-}+ 4\mathrm{j}$) and applying the same cuts as above, the authors in \cite{Biswal:2017nfl} find a value of $\sigma=2.2 \times 10^{-5} ~pb$. When multiplied by a $1 \%$ factor expected for electron charge misidentification at the ILC, they find this background is negligible.

Other possible backgrounds are SM events resulting in two genuine same-sign leptons, which could fake the same-sign dilepton signal, as backgrounds coming from the production of four on-shell $W$ bosons, with two like-sign ones decaying leptonically (with final neutrinos escaping undetected) and the other two decaying hadronically. For $\sqrt{s}=500 ~GeV$, this background can be estimated to be of order $10^{-5} ~pb$ adding the three possible final lepton flavors \cite{Peressutti:2011kx}. However, as these channels involve missing energy from the final neutrinos, they can be effectively suppressed by imposing appropriate cuts on the missing energy for the final states with muons and electrons ($l= e, \mu$)\cite{Zhang:2018rtr}. As an advantage over hadron colliders, the c.m. energy in lepton colliders is precisely measurable, and this helps the reconstruction of missing energy from the total energy-momentum unbalance in each event.  

The Majorana neutrino mass $m_N$ could be obtained in a reconstruction of the invariant mass of its decay products $M(l \mathrm{j} \mathrm{j})$, if the two final leptons (and the accompanying jets) can be isolated. This kind of reconstruction involves finding a resonant behavior of the invariant mass for these reconstructed objects \cite{Banerjee:2015gca}. The information on $m_N$, together with possible measurements of final state tau polarization can be used to give a hint on the kind of effective interactions taking part in the $N$ production and decay, as will be discussed in Sect.\ref{sec:tau_polarization}.

\section{Initial state polarization}\label{sec:ee_polarization}

The initial electron and positron polarizations can be used to distinguish the vectorial and scalar operators contribution to the studied process. The ILC is expected to operate in different polarization modes depending on the physics goals for each center of mass energy values. In particular, for $\sqrt{s}=500 ~GeV$ a running mode with opposite initial beam polarizations (H mode in \cite{Baer:2013cma}, table 1.1) is planned for increasing the luminosity in annihilation processes. In this section we consider three distinct initial polarization benchmark modes and test the ability to disentangle the vectorial and scalar operators contributions to the dominant double $N$ production process in Fig.\ref{fig:eeNN}.  
 
Under these conditions the relevant amplitude $|\overline{M_{I}}|^2$ in \eqref{eq:MI} can be written in terms of the initial electron ($P_{e^{-}}$) and positron ($P_{e^{+}}$) polarizations as 
\begin{eqnarray}\label{eq:inipol}
 |M^{e^{-} e^{+}}_{P_{e^{-}} P_{e^{+}}}|^2 = \frac{1}{4} (1-P_{e^{-}})(1+P_{e^{+}}) |M^{e^{-} e^{+}}_{LR}|^2 + \frac{1}{4} (1+P_{e^{-}})(1-P_{e^{+}}) |M^{e^{-} e^{+}}_{RL}|^2 
\end{eqnarray}
where the $LR$ and $RL$ amplitudes (left-polarized electron and right-polarized positron, and vice-versa) are 
\begin{eqnarray}
 |M^{{e^{-} e^{+}}}_{LR}|^2 &&= \frac{2}{\Lambda^4} (\alpha^{(1)}_{S_4}+ 2 \alpha_2)^2 (p_1.k_1) (p_2.k_2) 
  \nonumber
\\  
 |M^{{e^{-} e^{+}}}_{RL}|^2 &&= \frac{8}{\Lambda^4} \alpha^2_1 (p_1.k_2) (p_2.k_1) 
\end{eqnarray}
and $|M^{{e^{-} e^{+}}}_{LL}|=|M^{{e^{-} e^{+}}}_{RR}|=0$.

We find that while the amplitude for left-polarized electrons and right-polarized positrons gets contributions from both scalar and vectorial operators, the amplitude with right-polarized electrons and left-polarized positrons only receives vectorial contributions.

 \begin{figure*}[h]
\centering
\subfloat[]{\label{fig:pol_e0} \includegraphics[width=0.42\textwidth]{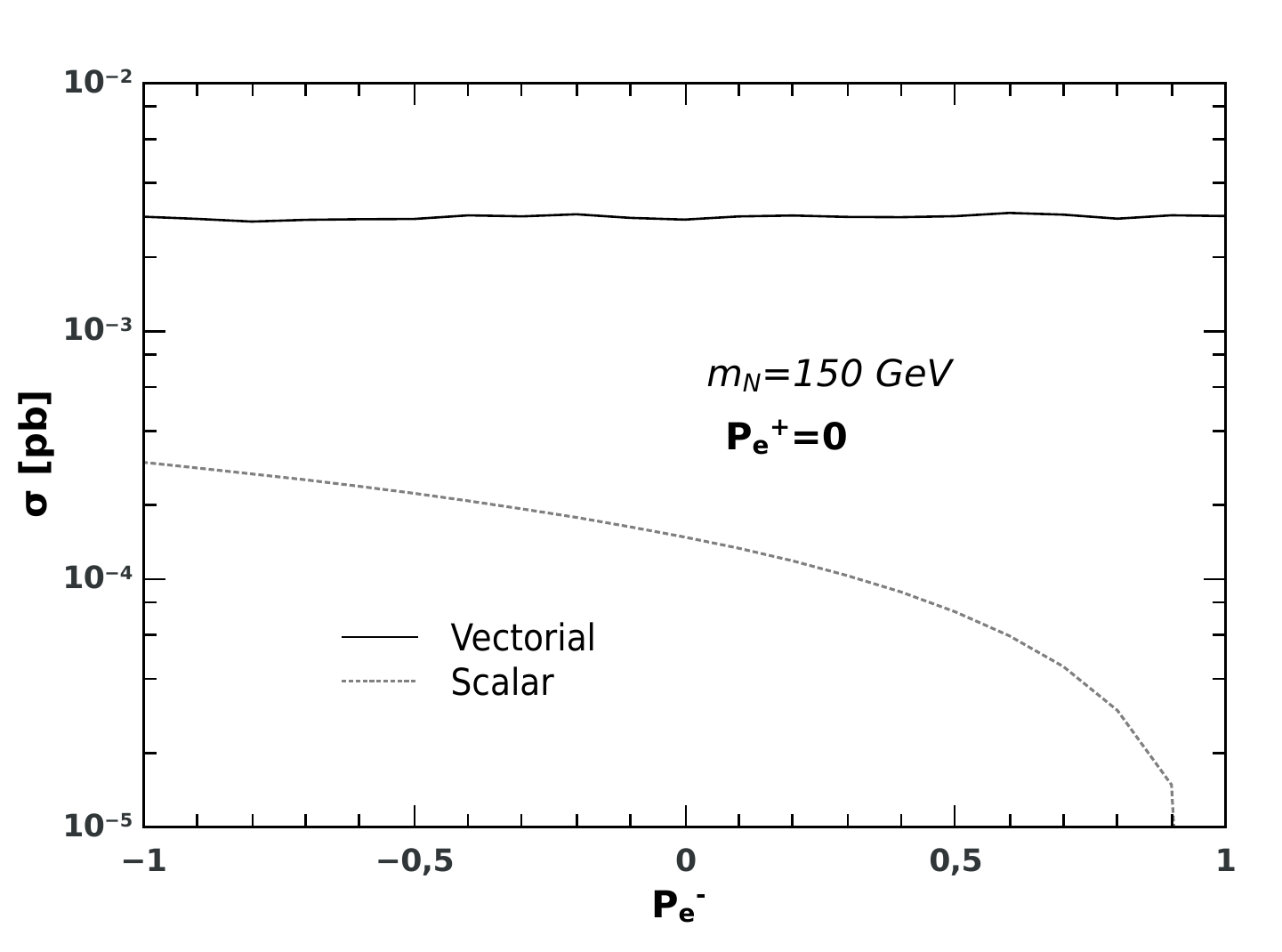}}

\subfloat[]{\label{fig:pol_eeq} \includegraphics[width=0.42\textwidth]{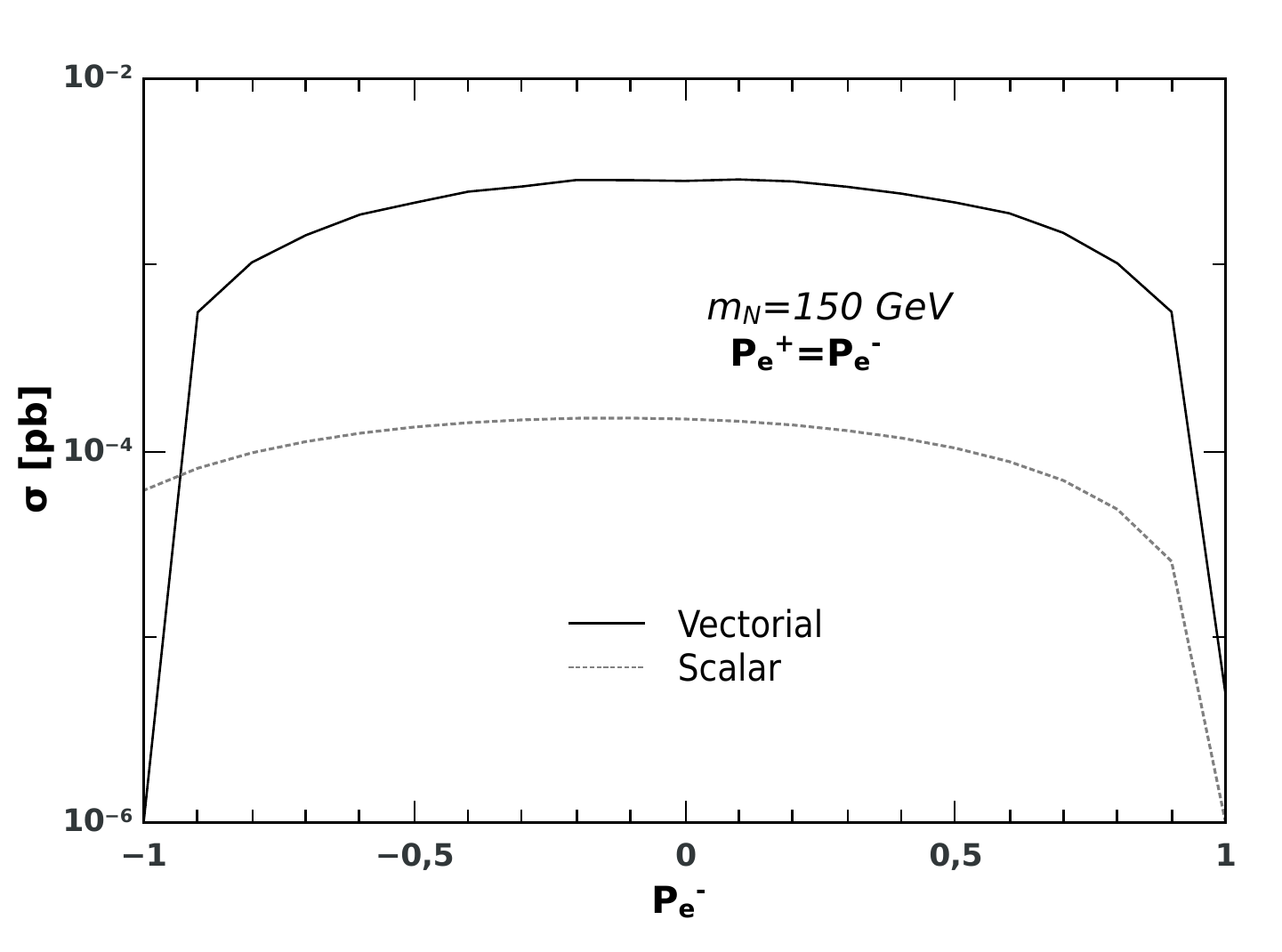}}

\subfloat[]{\label{fig:pol_eop} \includegraphics[width=0.42\textwidth]{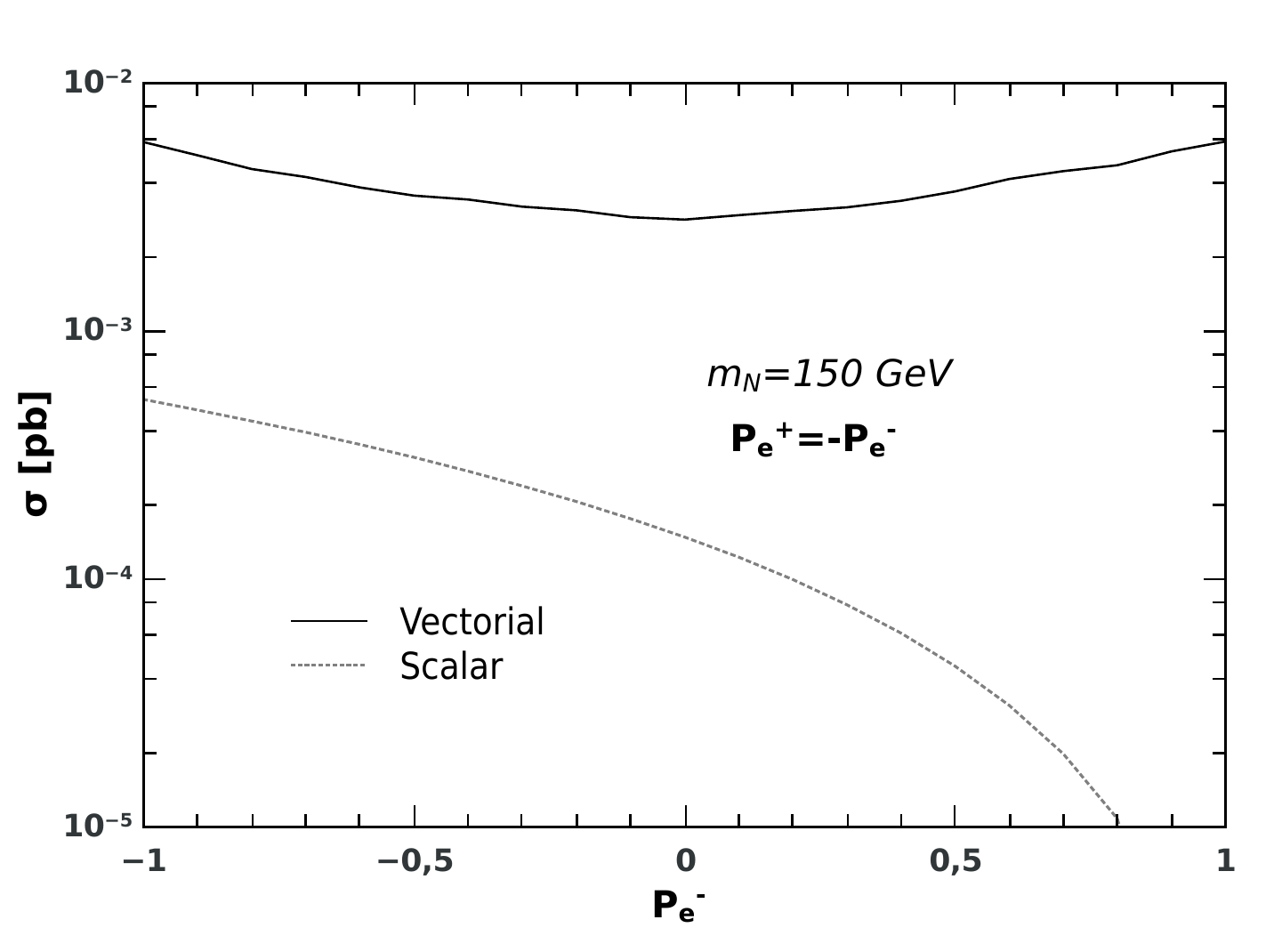}}
\caption{\label{fig:pol_ini}  Signal cross section as a function of the initial electron polarization.}
\end{figure*}

In Fig. \ref{fig:pol_ini} we show the contributions to the signal cross section for the process $e^{+}e^{-} \rightarrow N N \rightarrow l^{+} l^{+} + 4 \mathrm{j}$, with $l=e, \mu, \tau$ for Majorana neutrinos with mass $m_N=150 ~GeV$ given by vectorial and scalar operators, depending on the initial electron polarization $P_{e^{-}}$, for three different benchmark scenarios. In Fig. \ref{fig:pol_e0} the initial positron is taken to be unpolarized, in Fig. \ref{fig:pol_eeq} we take both initial polarizations to be the equal, and in Fig. \ref{fig:pol_eop} we take them to be opposite, as in the mentioned ILC H operation mode.

We find that for the unpolarized positron option (Fig. \ref{fig:pol_e0}) the vectorial operators give a cross section value in the $3 ~fb^{-1}$ range, mostly independent of the initial electron polarization value, while the scalars contribution decreases with positive $P_{e^{-}}$. For the equal polarization mode (Fig. \ref{fig:pol_eeq}) the two contributions have the same qualitative  behavior, despite the difference in magnitude. The opposite polarization mode (Fig. \ref{fig:pol_eop}) is the most promising to distinguish the kind of new physics contribution, as in this case the vectorial operators show a minimum contribution to the cross section when the initial beams are unpolarized, and the scalar operators contribution still decreases with positive $P_{e^{-}}$. Thus we find that comparing the cross section for different beam polarization configurations can help distinguish the possible vectorial or scalar effective interaction contributions. 

In the three plots we find that for $P_{e^{-}}=P_{e^{+}}=0$ (unpolarized beams), considering an integrated luminosity $\mathcal{L}=500 ~fb^{-1} $ it could be possible to separate the scalar and vectorial contributions up to a value of near $18$ standard deviations. In Fig. \ref{fig:pol_eop} we find that for $P_{e^{-}}=-P_{e^{+}}=0.8$ (opposite polarization beams) this number grows to $52$ sigma (see Eq.\eqref{eq:Spol}).

\begin{figure*}[h]
\centering
\includegraphics[width=0.45\textwidth]{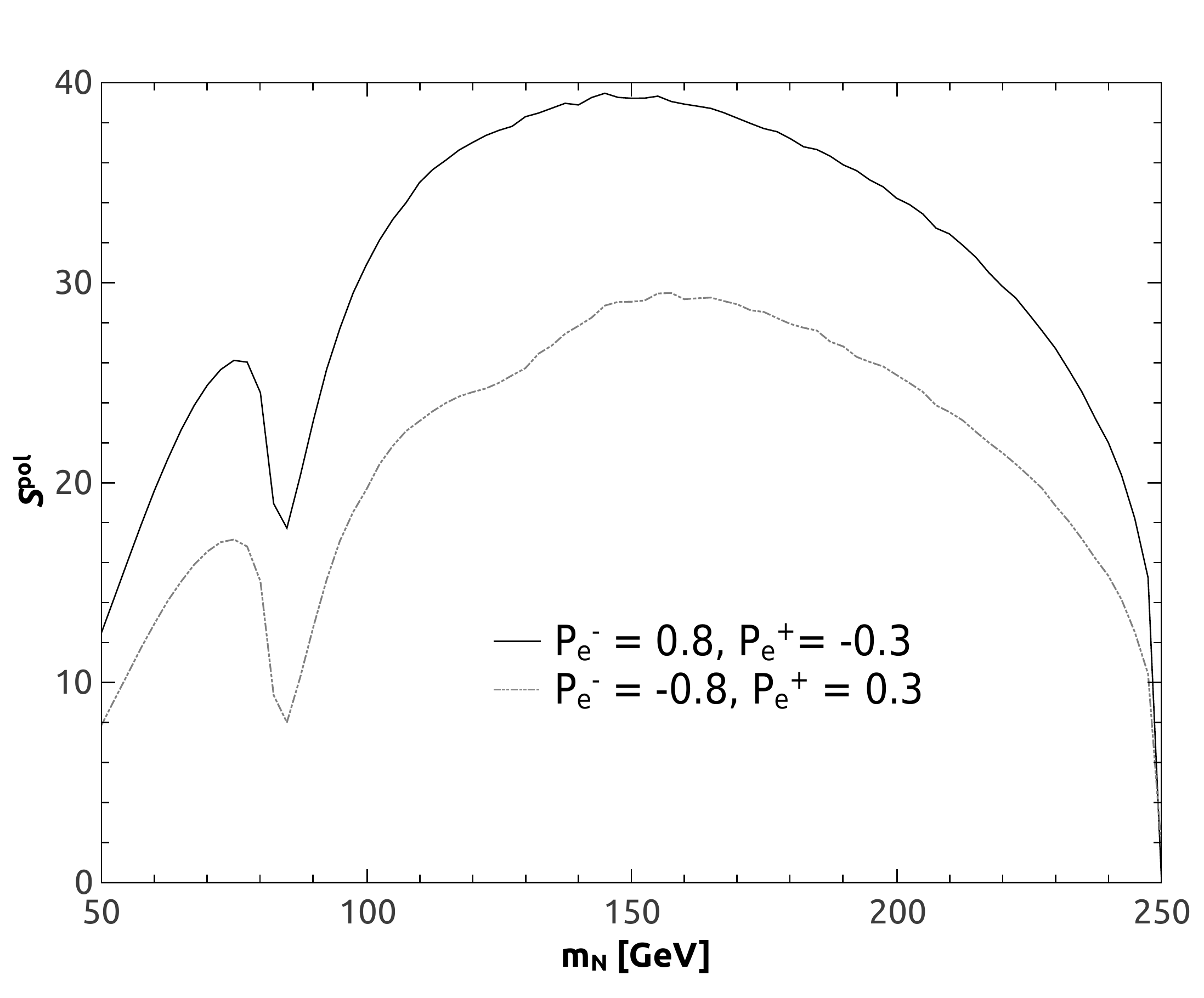}
\caption{\label{fig:Spol_mn} The polarization asymmetry $S^{pol}(m_N)$ for different initial beam polarizations.}
\end{figure*}

The cross section dependence  on $P_{e^{\pm}}$ in Eq. \eqref{eq:inipol} also allows us to compare the number of signal events produced by the vectorial and scalar effective interactions ($N^{vec}$ and $N^{sca}$ respectively) for different values of $P_{e^{-}}$ and $P_{e^{+}}$ when considering the production of Majorana neutrinos with different mass $m_N$. In order to explore the possibility of using polarized initial leptons to disentangle the contributions of the scalar and vectorial operators to the production cross section, we define the function $\mathcal{S}^{pol}$ as the number of standard deviations between the numbers of events produced by the vectorial and scalar operators contributions \cite{Duarte:2018xst}:
 \begin{equation}\label{eq:Spol}
 \mathcal{S}^{pol}=\frac{N^{vec}-N^{sca}}{\sqrt{N^{vec}}+\sqrt{N^{sca}}}
 \end{equation}
In Fig.\ref{fig:Spol_mn} we plot the values of the initial polarization asymmetry  $S^{pol}(m_N)$ for two possible fixed initial polarization settings ($P_{e^{-}}=0.8, P_{e^{+}}=-0.3$) and ($P_{e^{-}}= -0.8, P_{e^{+}}= 0.3$) \cite{Biswal:2017nfl}. We find that the two contributions could be very well separated in both beam operation modes, with the major difference arising in the right-polarized electron beam case. As an example, we find that for $m_N= 150 ~GeV$, taking a positive $P_{e^-}$ (solid line) the contributions from scalar and vectorial operators could be distinguished with a statistical significance of almost $40$ sigma, while for negative $P_{e^-}$ this value drops to $28$ sigma.

\section{Tau polarization signatures}\label{sec:tau_polarization}

Measurements of final state leptonic polarization have been crucial for the tests of the SM electroweak sector in lepton colliders. In particular, final tau and anti-tau polarization measurements at LEP and SLD experiments \cite{Decamp:1991vz, ALEPH:2005ab} have provided a direct measurement of the chiral asymmetries of the SM neutral current. Final taus are the only fermions whose polarization is accessible by means of the energy and angular
distribution of its decay products. These measurements rely on the dependence of kinematic distributions of the observed tau decay products on the helicity of the parent tau lepton. Recent studies at the LHC claim to have a statistical uncertainty comparable to similar measurements performed at LEP \cite{Cherepanov:2017afx}, and we expect improvements for the sensitivity in future detectors like the ILD at the ILC \cite{Jeans:2015vaa, Behnke:2013lya}.

The polarization of the final anti-taus can be used to distinguish the vectorial and scalar operators contributions. We define the leptonic final state polarization as
 \begin{equation}\label{eq:pol_taus}
 P_{\tau}=\frac{N_{++}+N_{+-}-N_{-+}-N_{--}}{N_{++}+N_{+-}+N_{-+}+N_{--}}
 \end{equation}
where the subscripts $+$ and $-$ in the number of events correspond respectively to Right and Left polarization states of the each final anti-tau $l_1$ and $l_2$ in Figs. \ref{fig:eeNN} and \ref{fig:eeNW}. Since the final anti-taus are identical particles, and the production and decay processes considered in the signal are the same for each of them, it is not possible to distinguish the final $+-$ and $-+$ polarization cases. So we expect that both numbers of events are equal: $N_{+-}=N_{-+}$, such that the polarization $P_{\tau}$ in \eqref{eq:pol_taus} is finally the ratio between the difference of the number of events for which both anti-taus are right-handed and left-handed, and the total number of events. 

In order to estimate the  error in the final state polarization $P_{\tau}$, we propagate it considering each number of events as Poisson distributed. Under these conditions we have
\begin{eqnarray}
\Delta P_{\tau}=\sqrt{\sum_{i,j=+,-}
\left(\frac{\partial P_{\tau}}{\partial N_{ij}} \right)^2 \left(\delta N_{ij}\right)^2}
\end{eqnarray}
where
\begin{equation}
\delta N_{++}=\sqrt{N_{++}} \; , \; \delta N_{+-}=\sqrt{N_{+-}} \; , \; \delta N_{-+}=\sqrt{N_{-+}} \; , \;
\delta N_{--}=\sqrt{N_{--}} .
\end{equation}
Thus we estimate the final state polarization error as 
\begin{equation} \label{eq:polerror}
\Delta P_{\tau} = \frac{2\left(N_{-+}+N_{--}\right)^{\frac12} \left(N_{++}+N_{+-}\right)^{\frac12}}
{\left( N_{++}+N_{+-}-N_{-+}-N_{--}\right)^{\frac32}}.
\end{equation}

To appreciate the ability of the final anti-taus polarization to determine the kind of effective operators involved in the studied interaction, we define a parameter  $\lambda \in [0,1]$ to measure the proportion of vectorial and scalar operators contributing to the process. Thus we multiply the vector operators by $\lambda$ and the scalars by $(1-\lambda)$, and study the dependence of the final polarization $P_{\tau}$ on this parameter for different Majorana neutrino masses $m_N$.  

As we found in the calculation of the Majorana neutrino decay $N \rightarrow \tau^{+} \mathrm{j} \mathrm{j}$ in Eqs. \eqref{eq:M_Ndec} and \eqref{eq:Ces}, the vectorial operators contribute to states with final Left anti-taus, and we expect to find a negative final polarization $P_{\tau}=-1$ for  a pure vectorial contribution ($\lambda=1$). Conversely, we expect a final $P_{\tau}=1$ for a pure scalar contribution ($\lambda=0$).   

\begin{figure*}[h]
\centering
\subfloat[]{\label{fig:pol_lan} \includegraphics[width=0.495\textwidth]{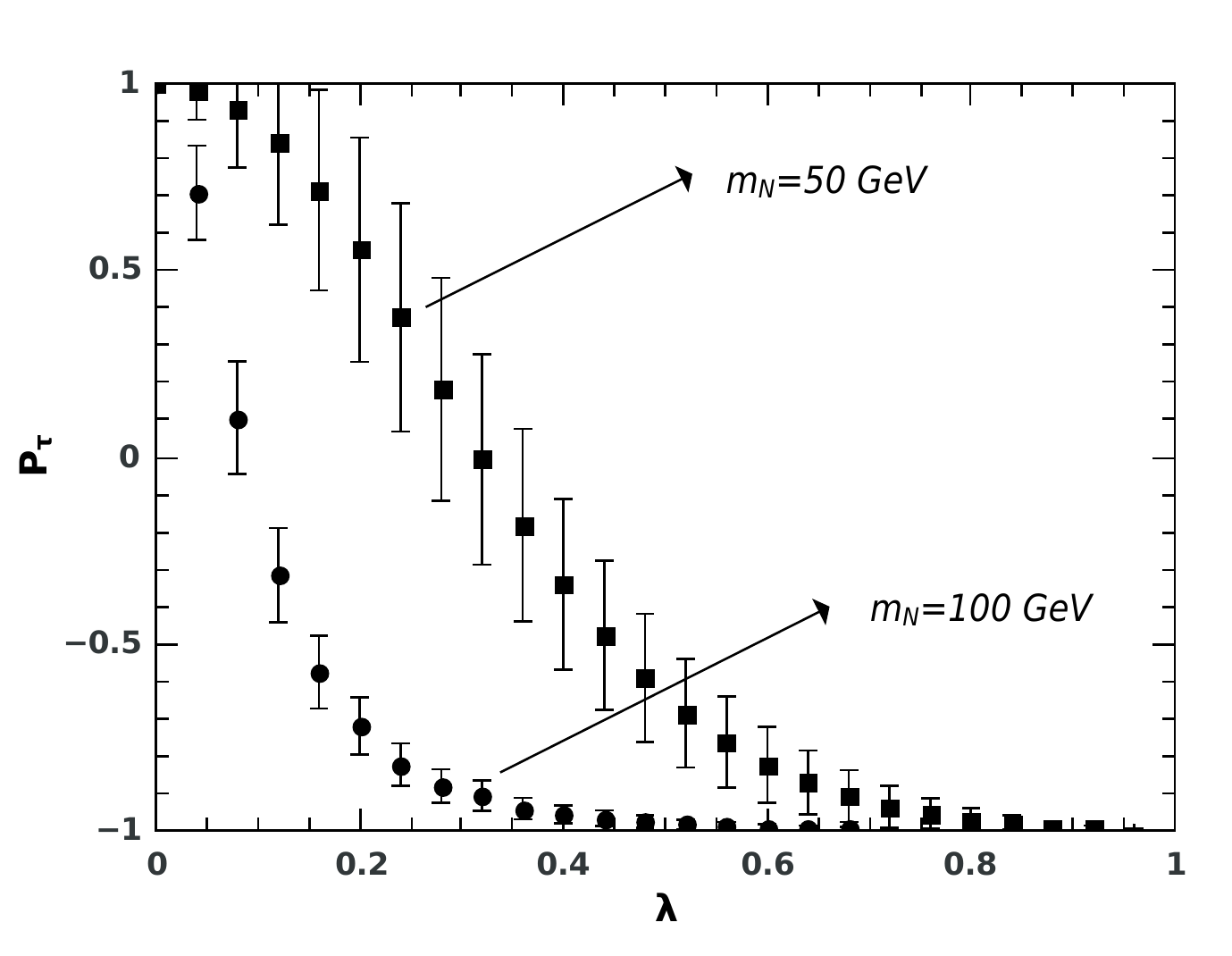}}~
\subfloat[]{\label{fig:pol_mn}  \includegraphics[width=0.505\textwidth]{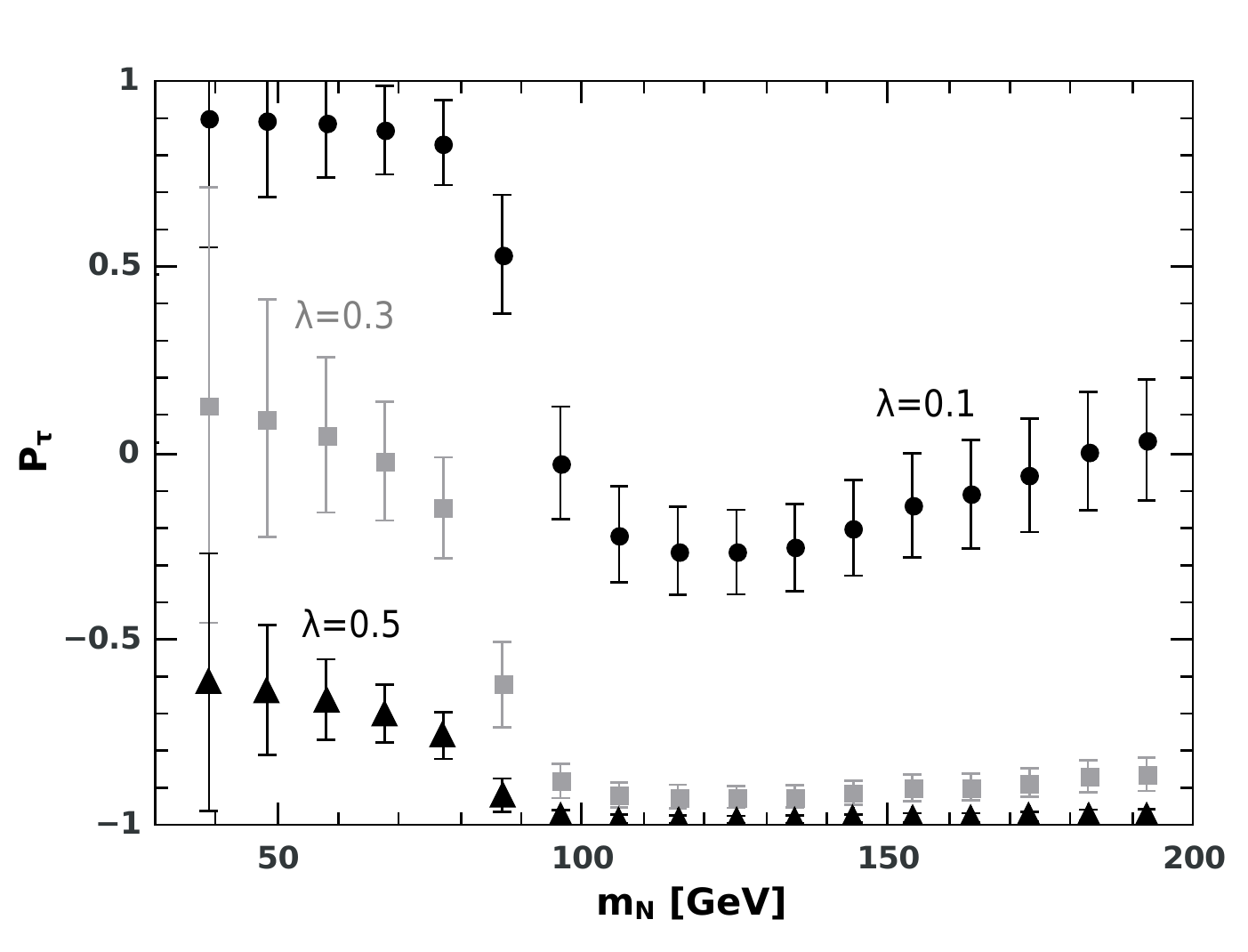}}
\caption{\label{fig:pol_tau} Final anti-tau polarization $P_{\tau}$ as defined in Eq.\eqref{eq:pol_taus} (a) as a function of $\lambda$ for different $m_N$ values and (b) as a function of $m_N$ for different $\lambda$ values.}
\end{figure*}

In Fig. \ref{fig:pol_tau} we plot the final state anti-tau polarization as a function of the variable $\lambda$ (Fig. \ref{fig:pol_lan}) and $m_N$ (Fig. \ref{fig:pol_mn}), respectively. In both figures we include the polarization errors, calculated as in Eq.\eqref{eq:polerror}, in order to appreciate the possibility of disentangling the kind of operators involved. 

In the case the studied LNV signal is detected and the Majorana neutrino mass $m_N$ is reconstructed, as we discussed in Sect. \ref{sec:numerical}, a measurement of the final state leptonic polarization $P_{\tau}$ could be able to determine the value of the parameter $\lambda$. For instance, by inspection of Fig. \ref{fig:pol_tau} one can see a positive final polarization $P_{\tau} \gtrsim 0$ for $m_N \approx 100 ~GeV$ would indicate the effective interaction to be mostly mediated by scalar operators.

\section{Summary and conclusions}\label{sec:Concl}

While models like the minimal seesaw mechanism lead to the decoupling of the heavy Majorana neutrinos, predicting mostly unobservable LNV, the effective Lagrangian framework considered in this work could serve as a means to discern between the different possible kinds of effective interactions contributing to LNV. The heavy neutrino effective field theory parameterizes high-scale weakly coupled physics beyond the minimal seesaw mechanism in a model independent framework, allowing for sizable LNV effects in colliders. In this work we investigate the $e^{+}e^{-} \rightarrow l^{+} l^{+} + 4 \mathrm{j}$ signal, mediated by Majorana neutrino effective interactions, which could be searched for in future lepton colliders \cite{Behnke:2013xla,dEnterria:2016sca,CEPCStudyGroup:2018rmc}.  

We have calculated the vectorial and scalar operators contribution to the signal cross section for different Majorana neutrino masses $m_N$, implementing basic trigger cuts for a benchmark ILC operating scenario with $\sqrt{s}=500 ~GeV$. In Sect.\ref{sec:ee_polarization} we calculate these contributions to the initially polarized cross section, for three different possible operation modes. We find that comparing the cross section dependence for different beam polarization configurations can help to the identification of the possible vectorial or scalar effective interactions contributions (Fig. \ref{fig:pol_ini}). 
We also define an initial polarization asymmetry $\mathcal{S}^{pol}$, which gives the number of standard deviations between the number of events produced by the vectorial-only or scalar-only interactions. Studying the dependence of this observable with the Majorana mass for two benchmark initial beam polarization configurations, we find the scalar and vectorial contributions could be well separated in both operation modes, with a greater difference in the case of a right polarized initial electron beam (Fig.\ref{fig:Spol_mn}). 

In Sect. \ref{sec:tau_polarization} we exploit the possibility to measure the final anti-taus polarization to study the chances to distinguish the vectorial and scalar contributions to the $e^{+}e^{-} \rightarrow N N \rightarrow \tau^{+} \tau^{+} + 4 \mathrm{j}$ signal. Weighting the vectorial and scalar operators by a factor $\lambda \in [0,1]$: with $\lambda=1$ (purely vectorial) and $\lambda=0$ (purely scalar) contributions (Fig. \ref{fig:pol_tau}) we find a measurement of the final polarization $P_{\tau}$ might be able to determine the value of the $\lambda$ parameter, provided that the mass $m_N$ can be reconstructed, possibly with the resonant invariant mass $M(\tau^{+} \mathrm{j} \mathrm{j})$ of its decay products.  

Our findings show that lepton colliders -where the clean environment allows for a detailed study of polarization observables- can provide relevant information on the kind of new physics responsible for lepton number violation in the $e^{+}e^{-} \rightarrow l^{+} l^{+} + 4 \mathrm{j}$ channel, complementing previous studies of LNV signals mediated by Majorana neutrinos with effective interactions at the LHC \cite{Duarte:2016caz} and in electron-proton colliders \cite{Duarte:2018xst, Azuelos:2018syu}. The initial beam and final tau polarization measurements could well disentangle possible vectorial and scalar operators contributions, which parameterize different high-scale physics beyond the minimal seesaw mechanism, giving us a hint on the possible physics contributing to (eventual) LNV, a fundamental puzzle in particle physics, as the nature of neutrino interactions.

{\bf Acknowledgements}

We thank CONICET (Argentina) and Universidad Nacional de Mar del
Plata (Argentina); and PEDECIBA and CSIC-UdelaR (Uruguay) for their 
financial supports.

\bibliographystyle{bibstyle.bst}
\bibliography{Bib_N_2_2019}

\end{document}